\documentclass[fleqn,10pt]{wlscirep}
\title{Schramm-Loewner evolution and perimeter of percolation clusters of correlated random landscapes}

\author[1,2,*]{C. P. de Castro}
\author[2]{M. Lukovi\'c}
\author[2]{G. Pompanin}
\author[1]{R. F. S. Andrade}
\author[2,3]{H. J. Herrmann}
\affil[1]{Instituto de F\'isica, Universidade Federal da Bahia, Campus Universit\'ario da Federa\c{c}\~ao, Salvador, BA,  40170-115, Brazil}
\affil[2]{ Computational Physics for Engineering Materials, IfB, ETH Zurich, Wolfgang-Pauli-Strasse 27, CH-8093 Zurich, Switzerland}
\affil[3]{Departamento de F\'isica, Universidade Federal do Cear\'a, Fortaleza, Cear\'a,60451-970, Brazil}
\affil[*]{ccastro@ethz.ch}

\begin{abstract}
 
Motivated by the fact that many physical landscapes are characterized  by  long-range  height-height  correlations that are quantified by the Hurst exponent $H$, we investigate the statistical properties of the iso-height lines of correlated surfaces in the framework of Schramm-Loewner evolution (SLE). We show numerically that in the continuum limit the external perimeter of a percolating cluster of correlated surfaces with $H \in \left[ -1,0 \right]$ is statistically equivalent to SLE curves. Our results suggest that
the external perimeter also retains the Markovian properties, confirmed by the absence of time correlations in the driving function and the fact that the latter is Gaussian distributed for any specific time. We also confirm that for all $H$ the variance of the winding angle grows logarithmically with size.

\end{abstract}
\begin{document}

\flushbottom
\maketitle
%
%
\thispagestyle{empty}

\section*{Introduction}
Random landscapes have been used as the basis for modeling a vast range of properties of different natural systems such as the sea surface temperature, ocean depth, height of land masses above sea level and plasma vorticity fields. Generally, such surfaces are correlated and in some cases they can have long range correlations that are characterized by the Hurst exponent, $H$. It has been shown recently that the iso-height lines taken at the percolation threshold of a long-range correlated random surface are scale-invariant with a fractal dimension $d_f$ that depends on $H$. What still remains elusive is whether these curves have a richer symmetry in the form of conformal invariance.

As in most physical systems, symmetry plays an important role in classifying and understanding the nature of these iso-height lines and the random landscapes from which they are extracted. For this reason, conformally invariant random curves extracted at critical heights of random surfaces and their fractal properties have received a lot of attention in the last decades \cite{Prakash1992StructuralPercolation,Kondev1995GeometricalSurfaces,KaldaPhysRevLett.90.118501,Schrenk2013PercolationDisorder}. The interest in such curves was triggered by the seminal works of Schramm \cite{Schramm2000,Lawler2001ValuesExponents}, who combined conformal mapping with stochastic processes into a process now known as Schramm-Loewner Evolution ($SLE$). $SLE$ is a one-parameter family of non-intersecting paths exhibiting conformal invariance that can be generated from Brownian motion whose diffusivity corresponds to the $SLE$ parameter, $\kappa$. 
It has been conjectured, and in a few cases proven, that $SLE_\kappa$ is the scaling limit of a variety of discrete random processes in two-dimensional space\cite{befara}. Inversely, it provides us with an alternative way to validate existing conjectures regarding the dependence of critical exponents on the Hurst exponent in percolation \cite{Schrenk2013PercolationDisorder,caio}. Moreover, the $SLE_\kappa$ approach allows us to generate directly such conformally invariant curves without the need to generating correlated surfaces or simulate growth models. Taking advantage of this feature, several numerical and empirical studies of correlated random systems such as turbulent vorticity fields\cite{Bernard2006ConformalTurbulence,Bernard2007InverseCurves}, graphene sheets \cite{Giordanelli2016ConformalSheets}, topology landscapes \cite{Boffetta2008HowShorelines}, percolation in correlated surfaces \cite{shortest} and accessible perimeters at fixed scale \cite{accessible}, have analyzed the corresponding two-dimensional random curves in the context of $SLE_{\kappa}$.

In what is known as \textit{chordal} $SLE$, 
the random continuous non-intersecting curve under study is parametrized over time such that at $t=0$ it starts from the origin located on the boundary of the upper half-plane $\mathbb{H}$ and tends to infinity as $t \rightarrow \infty$. Although such a curve does not intersect itself, in the continuum limit it might touch itself (although it still should not cross). 
The union of the space inside the loops formed when the trace touches itself, together with the curve up to time $t$ is called the hull and denoted by $\mathbb{K}_t$ \cite{cardy}. Such a definition guarantees a simple connected domain, i.e., a domain without holes, $\mathbb{H} \setminus \mathbb{K}$, bounded by the upper half plane. According to the Riemann mapping theorem, there exists an analytical function $g_t(z)$ which maps $\mathbb{H}\setminus \mathbb{K}_t$ into $\mathbb{H}$\cite{saberi}. This map satisfies the Loewner differential equation, 
\begin{equation}
\label{loewnerEq}
\frac{\partial g_t(z)}{\partial t} = \frac{2}{g_t(z) - \zeta_t}, \qquad \zeta_t = \sqrt[]{\kappa}B_t,
\end{equation}
where $g_0(z) = z$ and $\zeta_t$ is a continuum function called the \textit{driving function}. Schramm\cite{Schramm2000} proved that if the curves are conformal invariant and follow Markov properties, then $\zeta_t$ must be a Brownian motion with a single parameter $\kappa$. 

We present our study of iso-height lines of long-range correlated surfaces in the framework of chordal $SLE_{\kappa}$. More precisely, we investigated whether the complete perimeter of a percolating cluster obeys the $SLE_{\kappa}$ statistics properties in the continuum limit. By taking iso-height lines from correlated surfaces with  $-1 \leq H \leq 0$\cite{caio}, we find that the lines indeed do follow $SLE$ statistics. For $H=-1$ and $H=0$ we recover the analytical results that predict $\kappa = 6$ and $\kappa = 4$, respectively. Using the relationship between $\kappa$ and $d_f$ demonstrated by Beffara\cite{befara}, we also show that the conjectured $H$-dependencies of the diffusivity $\kappa$ and fractal dimension\cite{Schrenk2013PercolationDisorder,caio} $d_f$ mutually corroborate each other. Finally, we also verify the Markov property of the curves by showing that the corresponding driving functions are uncorrelated in time and that they follow Gaussian statistics.

\section*{Method}

We generate correlated random Gaussian surfaces on square lattices by associating to each lattice site ($x_{1}$,$x_{2}$) the height $h(\textbf{x}) = h(x_{1},x_{2})$ and we use the Fourier Filtering Method (FFM)\cite{Barnsley1988TheImages} in order to impose long-range correlations. Furthermore, we define the Hurst exponent associated with the correlation by choosing an appropriate power spectrum $S(\textbf{q})$ in the form of a power law such that,
\begin{equation}
\label{power}
S(\textbf{q}) \sim |\textbf{q}|^{-\beta_{c}} = \left( \sqrt{q_{1}^{2}+q_{2}^{2}} \right)^{-\beta_{c}} ,
\end{equation}
where $\beta_{c} = 2(H+1)$ \cite{Barnsley1988TheImages}. 
By multiplying a real-valued random variable $\textit{u}(\textbf{q})$ in two-dimensional Fourier space by the square-root of the power spectrum and subsequently applying the inverse Fourier transform, we obtain the correlated random Gaussian surface
\begin{equation}
h(\textbf{x}) = \Im^{-1} [ \sqrt{S(\textbf{q})} \textit{u}(\textbf{q}) ] .
\end{equation}
Without loss of generality, the two-dimensional random variable $\textit{u}(\textbf{q})$ is taken to be Gaussian distributed with unit variance.

According to the definition above, if $H=-1$ and therefore $\beta_{c} = 0$, the power spectrum in Eq. \ref{power} becomes independent of the frequency, giving rise to uncorrelated surfaces. As $H$ is increased from $-1$, height-height correlations are introduced into the surface. It should also be noted that, as a consequence of the extended Harris criterion  \cite{Dietrich1985IntroductionTheory,Smirnov2001CriticalPercolation,Schmittbuhl1993PercolationSurfaces,Sandler2004CorrelatedLevel,Weinrib1983CriticalDisorder,Janke2004Harris-LuckLattices}, there are some critical exponents of 2D systems that are not influenced by correlation effects introduced by $H \in [-1,-3/4]$, implying that for those Hurst values, the exponents are expected to be the same as for the uncorrelated system \cite{Schrenk2013PercolationDisorder}.

\subsection*{Winding Angle}
A simple and straight-forward necessary condition for conformal invariance is based on the statistical properties of the winding angle of the curve under study \cite{Boffetta2008HowShorelines}. Although the presence of conformal invariance does not guarantee $SLE$, it is certainly a necessary condition.
Since we are working with a discrete set of points that define the curve on the square lattice, we can consider the winding angle $\theta_i$ at a point $z_i$ to be the sum of all the turning angles $\alpha_i$ along the curve, starting from a point $z_0$\cite{Wieland2003contours}. Therefore, the winding angle at a point $z_N$ is given by
\begin{equation}\label{eq:windingAngleDef}
\theta_N = \sum_{i=0}^N\alpha_i,
\end{equation}
where $\alpha_i$ is the turning angle between two consecutive points on the curve. Curves that are conformally invariant have a probability distribution of the winding angle that is necessarily Gaussian with a variance that increases logarithmically with $L$ so that
\begin{equation}\label{eq:windingAngle}
Var[\theta_L]=\langle \theta_L^2 \rangle - \langle\theta_L\rangle^2 = a + m\ln L,
\end{equation}
where $a$ is a constant. Furthermore, it has been shown that for $SLE$ curves, $m=\kappa/4$ \cite{Schramm2000,Wieland2003contours,Duplantier1988}.

\subsection*{Driving Function - Direct SLE}

In order to determine whether a curve is indeed $SLE_\kappa$ and estimate the value of $\kappa$, we use the zipper algorithm with a vertical slit discretization \cite{Bauer,kennedy} to solve Eq. \ref{loewnerEq}. So,  given  a discrete curve in the upper half complex plane $(0,\gamma_1,...,\gamma_N)$, by using the inverse of $f_k(z)=g_k^{-1}(z)$\cite{kennedy}, its driving function can be recovered by applying the relations
\begin{equation}
t_k = \frac{1}{4} \sum_{i=1}^{k} Im \left\lbrace \omega_i \right\rbrace^2\quad\textrm{and}\quad\zeta_{t_k} = \sum_{i=1}^{k} Re \left\lbrace \omega_i \right\rbrace,
\label{timeDrive}
\end{equation}
where the $\omega_k$'s are determined recursively by
\begin{equation}
\omega_k = f_{k-1} \times f_{k-2} \times ... \times f_{1}(\gamma_k) , \qquad \omega_1 = \gamma_1
\end{equation}
and
\begin{equation}
f_k(z) = i\;\,\sqrt[]{-Im \left\lbrace \omega_i \right\rbrace^2 - \left( z - Re \left\lbrace \omega_i \right\rbrace  \right)^2}.
\end{equation}
Given that even for curves with equal length and step sizes the discretized times $t_k$ are not equally distributed, we linearly interpolate the measured driving function at equally spaced time intervals.

\section*{Results and Discussion}

\begin{figure}[ht!]
\centering
\includegraphics[scale=0.129]{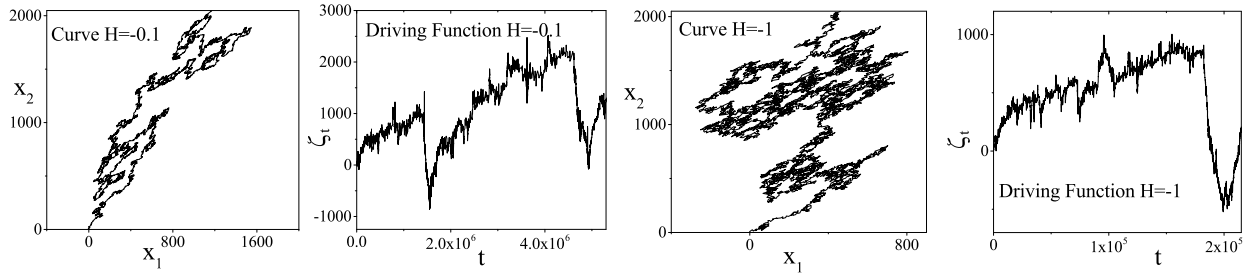}
\caption{Examples of full perimeters of percolating cluster for $H=-0.1$ and $H=-1$ and their respective driving function, calculated by the zipper algorithm. The jumps of the driving function reproduce the sinuosity of its respective curve. }
\label{curveAndDrive}
\end{figure}

\begin{figure}[ht!]
\centering
\includegraphics[scale=0.5,angle=-90]{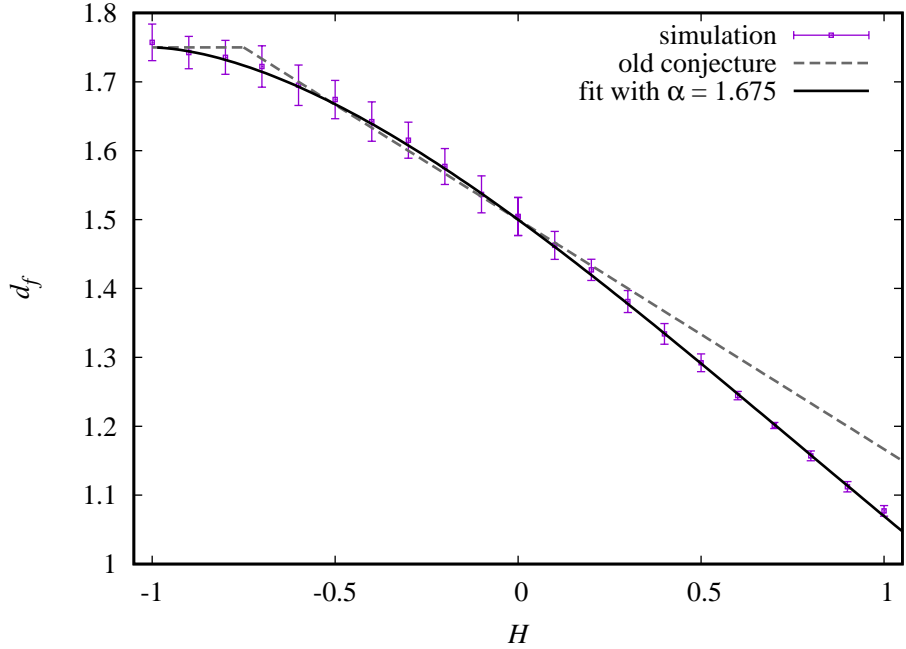}
\caption{Fractal dimension of the full perimeter as a function of $H$, calculated using the yardstick method\cite{caio}. The dashed gray line represents the old conjecture proposed by Schrenk \textit{et al.}\cite{Schrenk2013PercolationDisorder}. The black line corresponds to our fit using the model shown in Eq. \ref{eq:stretchedExp}, which is well supported also by the results of a numerical study by Castro \textit{et al.} \cite{caio} based on  different distributions of $\textit{u}(\textbf{q})$. The fit was done using all the numerical results obtained for $d_f$, beyond the interval $H\in[-1,0]$ considered in this report. All values are averages over $10^4$ samples and the error bars are defined by the variance of the distribution of the fractal dimension values.}
\label{perimeter}
\end{figure}

Our main goal is to study the properties and symmetries of the complete perimeter of the percolation cluster extracted from correlated landscapes with $H$ in the interval $[-1,0]$.
In Fig. \ref{curveAndDrive} we show examples of complete perimeters and their respective driving functions.
So far, analytical results for the critical exponents have been obtained only in the cases where $H=-1$ (uncorrelated surface) and $H=0$. Schrenk \textit{et al.} \cite{Schrenk2013PercolationDisorder} made the conjecture that the $H$-dependence of the complete perimeter fractal dimension has the form $d_f(H)=3/2-H/3$ for $H \in [-3/4,0]$. In this report, we present an alternative function that better fits the data that we obtained. We choose a compressed exponential function with the constraint imposed such that $d_f(-1)=7/4$ and $d_f(0)=3/2$. Consequently, only the power of the exponent $\alpha$ remains a free parameter to be chosen for the fit. We therefore obtain a simple analytical expression of the form: 
\begin{equation}\label{eq:stretchedExp}
d_f(H) = \frac{7}{4}\exp\left(-\ln(7/6)\cdot(H+1)^\alpha\right).
\end{equation}
It turns out that for $\alpha=1.675\pm0.003$ our model in Eq. $\ref{eq:stretchedExp}$ best fits the data in the interval $-1\leq H\leq 1$. In Fig. \ref{perimeter} we present our numerical results for the $H$-dependence of the interface fractal dimension, which support the choice of function used for the fit.  Moreover, the $H$-dependence was later also shown to be independent of the shape of the distribution of the random numbers, $\textit{u}(\textbf{q})$, used to generate the correlated landscapes \cite{caio}.

It was conjectured by Rohde and Schramm\cite{rohde} and demonstrated by Beffara\cite{befara} that the $SLE_\kappa$ curves are fractals whose dimension, $d_f$, is related to the diffusion coefficient, $\kappa$, by the expression
\begin{equation}
d_f = min \left( 1 + \frac{\kappa}{8},2   \right).
\label{beffaraEq}
\end{equation}
Therefore, the accuracy of the value of $\kappa$ estimated from a random curve can be verified by comparing the value of $d_f$ obtained via Eq. \ref{beffaraEq} with the value of $d_f$ determined directly using scale invariant methods such as the yardstick method.

\begin{figure}[ht]
\centering
\hspace{-0.8cm}\includegraphics[scale=0.35,angle=-90]{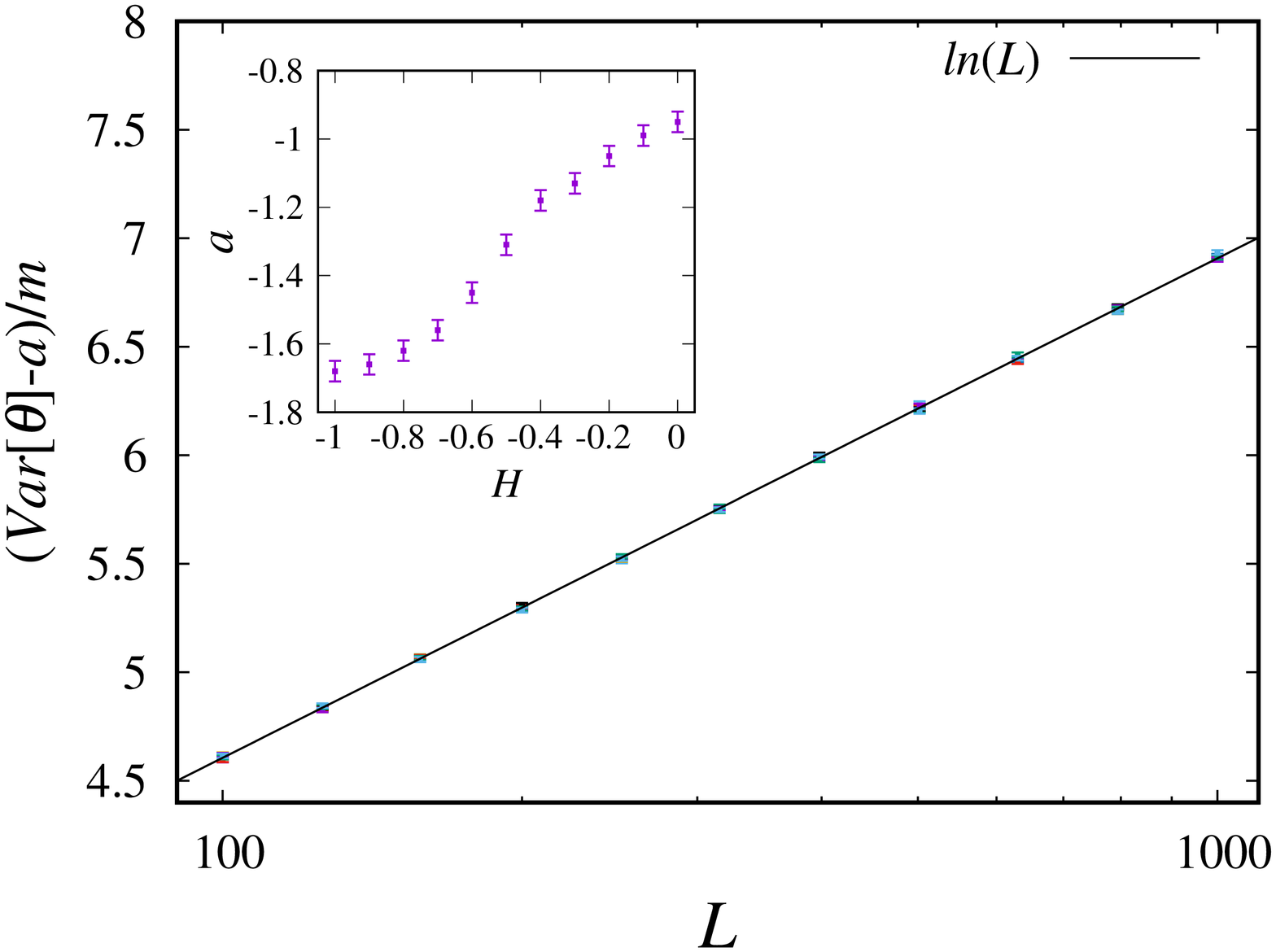}\hspace{-0.2cm}
\includegraphics[scale=0.35,angle=-90]{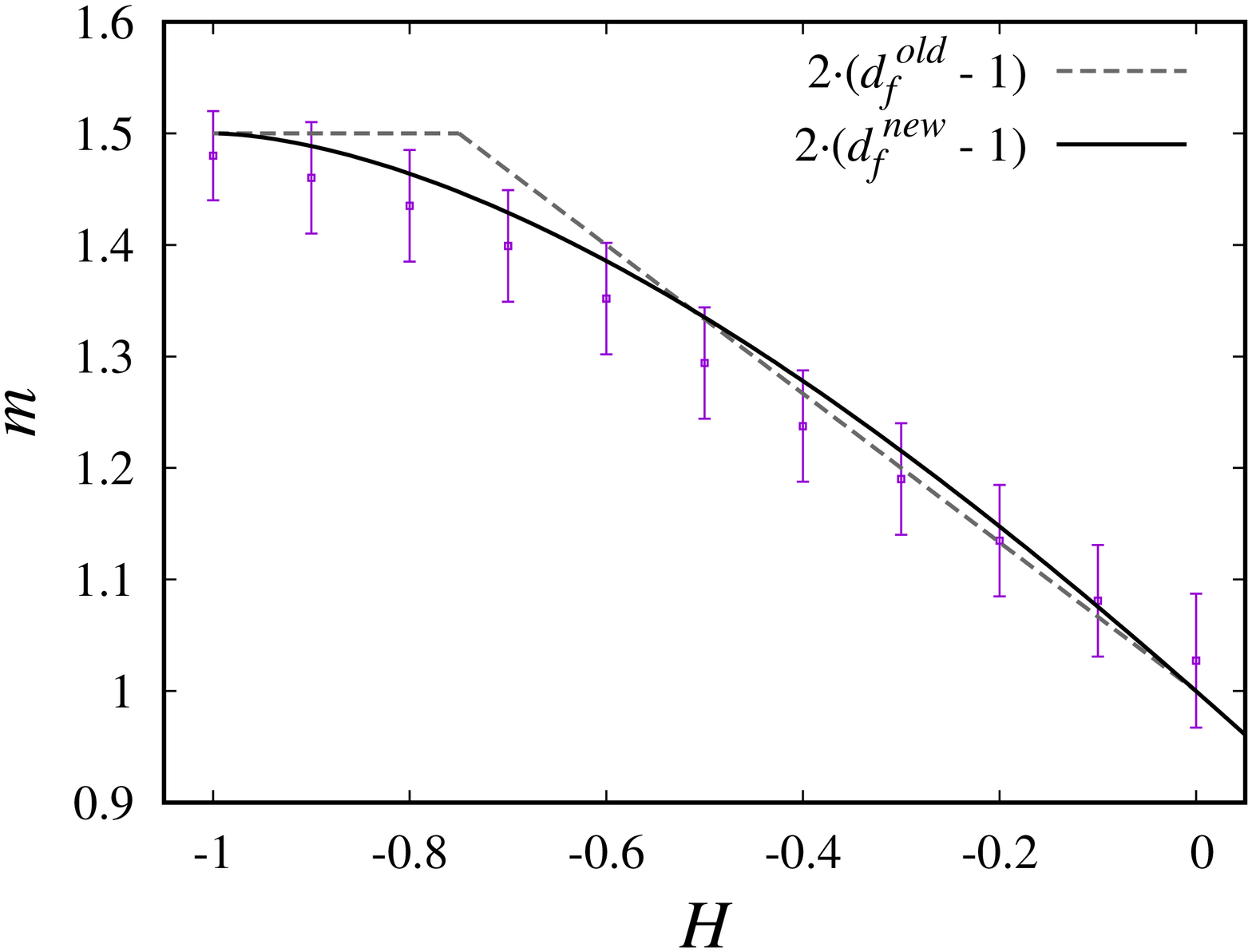}

\caption{\textbf{Left (main):} Rescaled variance of the winding angle distribution as a function of system size for different values of the Hurst exponent. The points (error bars are only slightly larger than the symbols) correspond to values of $H$ in the interval $[-1,0]$ with increments of 0.1. The results confirm the relationship in Eq. 5 and therefore support the presence of conformal invariance in the complete perimeter of percolating clusters considered in this study. \textbf{Left (insert):} Values obtained numerically for the constant $a$ in Eq. \ref{eq:windingAngle} as a function of $H$. \textbf{Right:} The data points correspond to the slope $m$ in Eq. \ref{eq:windingAngle}. The black curve is derived from the fit of the fractal dimension, $d_f$, in Eq. \ref{eq:stretchedExp} with $\alpha=1.675$. The dashed gray curve is derived from the old conjecture of the fractal dimension.}
\label{fig:windingAngle}
\end{figure}

For the complete perimeters we calculated the variance of the distribution of all the winding angles $\theta$ with respect to the origin of the curve in a lattice of size $L$. We determined the winding angle at each point $z_N$ of the perimeter according to the definition in Eq. \ref{eq:windingAngleDef} and then calculated the variance of the resulting distribution. In Fig. \ref{fig:windingAngle} we present our numerical results for different lattice sizes $L$ and different values of $H$. We show that for all values of $H\in[-1,0]$ considered, the variance does indeed grow logarithmically with system size. The expression in Eq. \ref{eq:windingAngle} fits all our data, which is one condition for the curves to be conformally invariant. By making the assumption, for which later we will give support, that the curve is also $SLE$ and that therefore $m=\kappa/4=2(d_f-1)$, in Fig. \ref{fig:windingAngle} we compare our results obtained for the values of $m$ with the fit made using Eq. \ref{eq:stretchedExp}. 


\begin{figure}[!ht]
\centering
\includegraphics[scale=0.32]{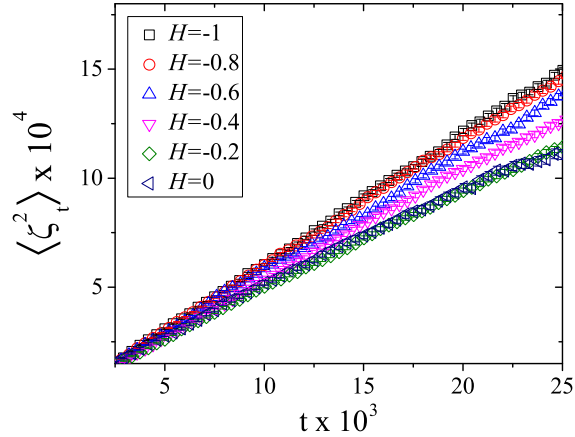}
\caption{The linear time dependence of the mean square displacement of the driving function for different values of Hurst exponent. Without loss of information, in this plot we did not show all the data points used for the full calculation of the $\kappa$. All values are average over $10^4$ samples and the error bars (inside the symbols) are defined by the variance of the mean square displacement distribution of the driving function.}
\label{variAllH}
\end{figure}

Given that the winding angle test alone is not sufficient to determine whether a curve is $SLE$, we focus on the direct approach and study the properties of the driving function of the complete perimeter. In the case where the random curve is $SLE$ in the scaling limit, the resulting driving function is a Brownian motion with mean square displacement that scales with time as
\begin{equation}
\left\langle   \zeta_t^2 \right\rangle \sim \kappa t.
\label{linear}
\end{equation}
We therefore investigate this dependence for critical site percolation interfaces of random landscapes with $H$ values in the interval $[-1,0]$. As shown in Fig. \ref{variAllH}, we obtain a good linear dependence  of the variance on time. The different slopes ($\kappa$ values)  are due the different Hurst exponents of the random surfaces from which the curves were extracted. The mean square displacement error $\Delta \left\langle \zeta_t^2 \right\rangle $ was computed as follows:
\begin{equation}
\Delta \left\langle   \zeta_t^2 \right\rangle = \sqrt[]{\frac{1}{N} \left[ \left\langle   \zeta_t^4 \right\rangle -  \left\langle   \zeta_t^2 \right\rangle^2  \right]}, \qquad \left\langle   \zeta_t^4 \right\rangle = \frac{1}{N} \sum_{k=1}^{N}  \zeta_t^4,
\end{equation}
where $N$ is the total number of samples of driving functions. 

\begin{figure}[!ht]
\centering
\includegraphics[scale=0.12]{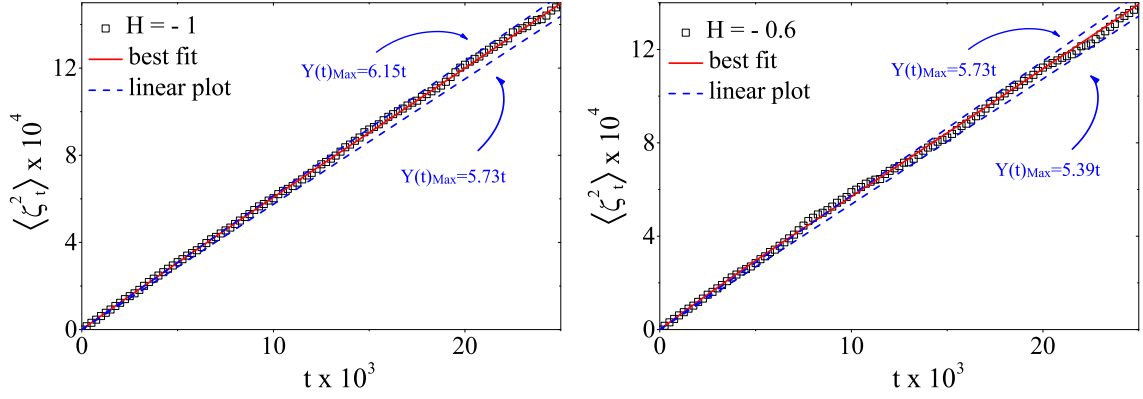}
\caption{ Linear time dependence of the mean square displacement of the driving function for two different Hurst exponents. The red line corresponds to the best fit using the model in Eq. \ref{newFit}. The dashed blue lines $Y_{Max}(t)$ and $Y_{Min}(t)$ are linear functions with maximum and minimum slopes ($\kappa$ values) which delimit the time evolution of $\left\langle   \zeta_t^2 \right\rangle$. These limits were used to define the error bars of the numerical estimate of $\kappa$, as shown in the Fig.\ref{KappaH}.} 
\label{variKappa}
\end{figure}

\begin{figure}[!ht]
\centering
\includegraphics[scale=0.35]{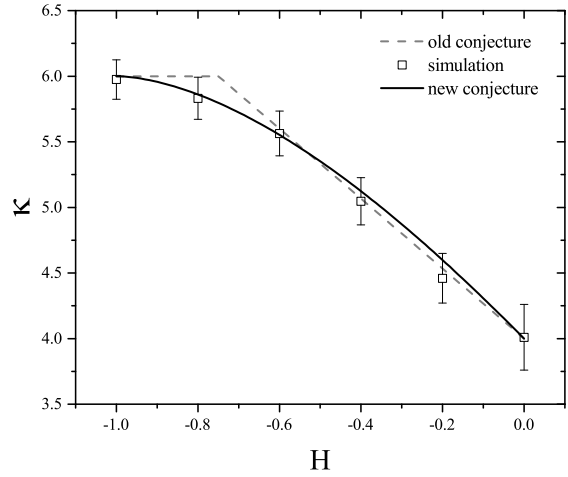}
\caption{Dependence of the diffusion coefficient ($\kappa$) on the Hurst exponent ($H$) estimated by Eq. \ref{kappaDefi}. The values of the points were determined via the direct SLE test. The dashed gray line (old conjecture) was  derived by combining Eq. \ref{beffaraEq} with the conjecture put forward by Schrenk \textit{et al.}\cite{Schrenk2013PercolationDisorder}, ($d_f=\frac{3}{2}-\frac{H}{3}$). The black line was obtained using the new fit in  Eq.  \ref{eq:stretchedExp}.}
\label{KappaH}
\end{figure}

In order to determine the value of $\kappa$ we used the function,
\begin{equation}
f(t)=\kappa_{lim} t + \frac{t^d}{ct^e+b},\qquad e>d>0,\quad b>c>0,\quad b,c,d,e\in\mathbb{R} \quad .
\label{newFit}
\end{equation}
as our model to fit the data of the time evolution of $\left\langle\zeta_t^2 \right\rangle$  (Fig. \ref{variKappa}). In the limit $t \rightarrow \infty$,  the linear term prevails due to the condition $d>e$,  being the non linear term only relevant in the low $t$ range.
We then considered two straight lines $Y(t)_{Max}$ and $Y(t)_{Min}$, that bound  the evolution of $\left\langle\zeta_t^2 \right\rangle$, estimating then the maximum and minimum values of $\kappa$ , respectively (see Fig.\ref{variKappa}). Finally, we calculated $\kappa$ and its corresponding error with the following expression:
\begin{equation}
\kappa = \kappa_{lim} \pm \left( \frac{\kappa_{max} - \kappa_{min}}{2}   \right).
\label{kappaDefi}
\end{equation}
Following Eq. \ref{kappaDefi}, we calculated $\kappa$ for a family of curves associated with different values of $H$. In Fig. \ref{KappaH} we compare the values of $\kappa$ calculated numerically with the two conjectures mentioned earlier in the text. As we can see, the old conjecture\cite{Schrenk2013PercolationDisorder} does not agree with the numerical estimation for the complete range of $\kappa$ values. On the other hand, the numerical estimates of $\kappa$ fit with the new fit which was derived from Eq. \ref{eq:stretchedExp} and Eq. \ref{beffaraEq}.

\begin{figure}[!ht]
\centering
\includegraphics[scale=0.35]{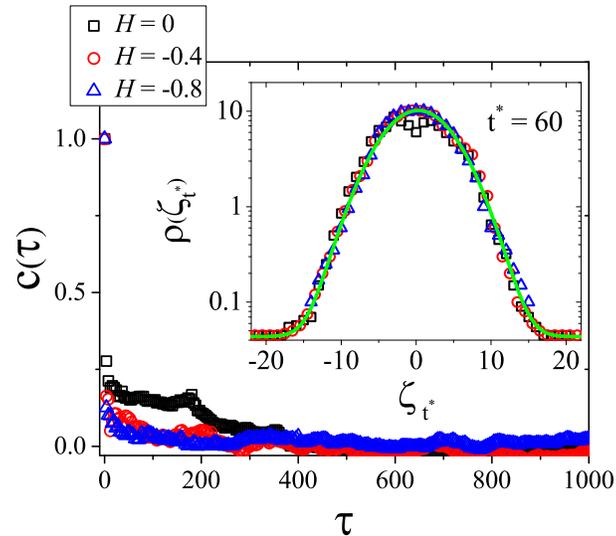}
\caption{Correlation time of the driving function for three different values of Hurst, $H=-0.8, -0.4, 0$.The inset shows the probability density distribution $\rho(\zeta)$ for a specific Loewner time $t^*=60$ for the same Hurst values described above. The solid green line is a guide to the eye $\rho(\zeta) = \frac{1}{\sqrt[]{2\pi \kappa t^*}} \exp \left( \frac{- \zeta_{t^*}^{2}}{2\kappa t^*} \right)$.}
\label{corretime}
\end{figure}

In order to confirm that a random curve is $SLE$ it is not sufficient that the evolution of the mean square displacement of the corresponding driving function is linear in time as shown in Fig. \ref{variKappa}. It is also necessary that the driving function is uncorrelated in time. We therefore tested for the Markov property of the driving function by computing its time correlation function $c(t,\tau)$, defined by:
\begin{equation}
c(t,\tau) = \frac{\left\langle \zeta_{t+\tau}\zeta_{t}   \right\rangle - \left\langle \zeta_{t+\tau} \right\rangle \left\langle \zeta_{t} \right\rangle}{\sqrt[]{\left( \left\langle \zeta_{t+\tau}^2  \right\rangle - \left\langle \zeta_{t+\tau}  \right\rangle^2  \right)\left( \left\langle \zeta_{t}^2  \right\rangle - \left\langle \zeta_{t}  \right\rangle^2  \right)}}.
\end{equation}
As shown in Fig. \ref{corretime} the correlation $c(t,\tau)$ goes to zero after a few times steps, as expected for Brownian motion. The short time correlation is associated to the discretization of the curve, i.e. due to the finite grid size. To complete the investigation of the Markov property we also calculated the distribution of the driving function for a specific time ($t^*$), which is shown to follow a Gaussian distribution (see inset of Fig. \ref{corretime}).

\section*{Conclusion}

Given that many systems can be viewed as long-range correlated landscapes, properties of the iso-height lines extracted from them become relevant. Our results suggest that the complete perimeter of the percolating cluster of long-range correlated landscapes ($-1 \leq H \leq 0$) is statistically equivalent to $SLE$ curves. We found consistent agreements between the diffusion constant $\kappa$ calculated by the zipper algorithm and the value obtained via the fractal dimension of the $SLE$ curves\cite{befara}. We also proposed a new conjecture for the dependence between $\kappa$ and $H$, in the assumed interval, on correlated random surfaces.  In addition, we also showed that, in the scaling limit, the curves are Markovian in nature, in the sense that their driving functions are uncorrelated in time and Gaussian distributed at specific points in time. A practical consequence of having established that the curves under study are $SLE$ is that we can extend the established results from $SLE$ to iso-height lines. Indeed, it is possible to generate an ensemble of such curves just by solving a stochastic differential  equation,  without the need to generate the entire landscape.

\section*{Acknowledgements}

We acknowledge the financial support from European Research Council (ERC) Advanced Grant 319968 FlowCCS, the ETH Risk Center, the Brazilian INCT-SC, and Minist\'erio da Educa\c{c}\~ao do Brasil (Funda\c{c}\~ao CAPES).

\section*{Author contributions statement}

C. P. de Castro, M. Lukovi\'c, R. F. S. Andrade and H. J. Herrmann conceived the research, C. P. de Castro conducted the numerical simulations. G. Pompanin contributed to the simulations and results relative to the winding angle. All authors contributed to the writing of the manuscript.
\section*{Additional information}

\textbf{Competing financial interests:}
The authors declare no competing financial interests.

\end{document}